\documentclass[letterpaper,11pt]{article}

\usepackage[margin=1.0in]{geometry}

\newcommand{\Id}[1]{{\texttt{#1}}}

\usepackage{hyperref}

\usepackage{textcomp}
\usepackage{amsmath}
\usepackage{amsfonts}
\usepackage{amsthm}
\usepackage{algorithm}
\usepackage[noend]{algpseudocode}
\usepackage{mathtools}
\usepackage{wasysym}
\usepackage{array}
\usepackage{subcaption,siunitx,booktabs}
\usepackage{numprint}
\npdecimalsign{.} 

\usepackage{color, colortbl}
\definecolor{Gray}{gray}{0.9}
\definecolor{darkblue}{RGB}{0, 0, 255}

\DeclarePairedDelimiter{\ceil}{\lceil}{\rceil}
\usepackage{xspace}

\newcommand{\erdos}{Erd{\H o}s-R{\'e}nyi}

\def\MdR{\ensuremath{\mathbb{R}}}

\newcommand{\email}[1]{\href{mailto:#1}{#1}}

\title{Engineering Kernelization for Maximum Cut\footnote{The research leading to these results has received funding from the European Research Council under the European Union's Seventh Framework Programme (FP/2007-2013) / ERC Grant Agreement no. 340506.}}

\author{Damir Ferizovic\footnote{Karlsruhe Institute of Technology, Karlsruhe, Germany, \email{dferizovic@ira.uni-karlsruhe.de}}
\and
Demian Hespe\footnote{Karlsruhe Institute of Technology, Karlsruhe, Germany, \email{hespe@kit.edu}}
\and
Sebastian Lamm\footnote{Karlsruhe Institute of Technology, Karlsruhe, Germany, \email{lamm@kit.edu}}
\and
Matthias Mnich\footnote{Universit{\"a}t Bonn, Bonn, Germany, \email{mmnich@uni-bonn.de}, \emph{Supported by DFG grant MN 59/1-1.}}
\and
Christian Schulz\footnote{University of Vienna, Faculty of Computer Science, Vienna, Austria, \email{christian.schulz@univie.ac.at}}
\and
Darren Strash\footnote{Hamilton College, Clinton, New York, USA, \email{dstrash@hamilton.edu}}}

\makeatletter
\newcommand{\manuallabel}[1]{\def\@currentlabel{#1}}
\makeatother

\newcounter{KernelizationRule}
\newenvironment{KernelizationRule}[1][]{\refstepcounter{KernelizationRule}\par\medskip
	\noindent \textbf{Reduction Rule~\theKernelizationRule. #1} \rmfamily}{\smallskip}

\newcounter{KernelizationRuleA}

\newcounter{OneWayRule}

\newenvironment{ExactKernelizationRule}[1]{\manuallabel{#1}\par\medskip
	\noindent \textbf{Reduction Rule~#1.} \rmfamily}{\smallskip}

\newcommand{\mycite}[1]{\cite{#1}}
\newcommand{\citeauthor}[1]{\cite{#1}}
\newcommand{\unaryminus}{\protect\scalebox{0.75}[1.0]{$-$}}
\newcommand{\unaryplus}{\protect\scalebox{1.0}[1.0]{$+$}}

\begin{document}

\date{}
\maketitle

\begin{abstract}
  Kernelization is a general theoretical framework for preprocessing instances of $\mathsf{NP}$-hard problems into (generally smaller) instances with bounded size, via the repeated application of data reduction rules.
  For the fundamental {\sc Max Cut} problem, kernelization algorithms are theoretically highly efficient for various parameterizations.
  However, the efficacy of these reduction rules in practice---to aid solving highly challenging benchmark instances to optimality---remains entirely unexplored.

We engineer a new suite of efficient data reduction rules that subsume most of the previously published rules, and demonstrate their significant impact on benchmark data sets, including synthetic instances, and data sets from the VLSI and image segmentation application domains.
Our experiments reveal that current state-of-the-art solvers can be sped up by up to \emph{multiple orders of magnitude} when combined with our data reduction rules.
On social and biological networks in particular, kernelization enables us to solve four instances that were previously unsolved in a ten-hour time limit with state-of-the-art solvers; three of these instances are now solved in less than two seconds.

\end{abstract}

\section{Introduction}
\label{sec:Introduction}
The (unweighted) {\sc Max Cut} problem is to partition the vertex set of a given graph {$G=(V,E)$} into two sets {$S\subseteq V$ and $V \setminus S$} so as to maximize the total number of edges between those two sets. Such a partition is called a \emph{maximum cut}. Computing a maximum cut of a graph is a well-known problem in the area of computer science; it is one of Karp's 21 $\mathsf{NP}$-complete problems~\cite{karp1972reducibility}
While signed and weighted variants are often considered throughout the literature~\cite{barahona1982computational,barahona1996network,barahona1988application,chiang2007fast,de2013estimation,harary1959measurement,harary2002signed}, the simpler (unweighted) case still presents a significant challenge for researchers, and solving it quickly is of paramount importance to all variants. 
{\sc Max Cut} variants have many applications,
including social network modeling~\cite{harary1959measurement}, statistical physics~\cite{barahona1982computational}, portfolio risk analysis~\cite{harary2002signed}, VLSI design~\cite{barahona1988application,chiang2007fast},
network design~\cite{barahona1996network}, and image segmentation~\cite{de2013estimation}.

Theoretical approaches to solving {\sc Max Cut} {primarily} focus on producing efficient parameterized algorithms through \emph{data reduction rules}, which reduce the input size in polynomial time while maintaining the ability to compute an optimal solution to the original input. 
If the resulting (irreducible) graph has size bounded by a function of a given parameter, then it is called a \emph{kernel}.
Recent works focus on parameters measuring the distance $k$ between the maximum cut size of the input graph and a lower bound $\ell$ guaranteed for all graphs.
The algorithm then must decide if the input graph admits a cut of size~$\ell+k$ for a given integer $k\in\mathbb N$.
Two such lower bounds are the Edwards-Erd{\H{o}}s bound~\cite{edwards1973some,edwards1975improved} and the spanning tree bound.
{Crowston et al.~\cite{DBLP:journals/corr/abs-1112-3506} were the first to show that unweighted {\sc Max Cut} is fixed-parameter tractable when parameterized by distance $k$ above the Edwards-Erd{\H{o}}s bound.
Moreover, they show the problem admits a polynomial-size kernel with $O(k^{5})$ vertices.
Their result was extended to the more general \textsc{Signed Max Cut} problem, and the kernel size was decreased to $O(k^{3})$ vertices~\cite{DBLP:journals/corr/abs-1212-6848}.
Finally, Etscheid and Mnich~\cite{Etscheid2018} improved the kernel size to an optimal $O(k)$ vertices even for signed graphs, and showed how to compute it in linear time $O(k \cdot (|V| + |E|))$.

Many practical approaches exist to compute a maximum cut or (alternatively) a large cut.
Two state-of-the-art exact solvers are \textsc{Biq Mac} (a solver for \emph{bi}nary \emph{q}uadratic and \emph{Ma}x-\emph{C}ut problems) by Rendl et al.~\cite{RRW10}, and \textsc{LocalSolver}~\cite{localsolverx1,localsolverx2}, a powerful generic local search solver that also verifies optimality of a cut. 
Many heuristic (inexact) solvers are also available, including those using unconstrained binary quadratic optimization~\cite{Wang2013}, local search~\cite{Benlic2013}, tabu search~\cite{Kochenberger2013}, and simulated annealing~\cite{Arraiz2009}. 

Curiously, data reduction, which has shown promise at preprocessing large instances of other fundamental $\mathsf{NP}$-hard problems~\cite{abu2007kernelization,hespe2017scalable,lamm2017finding}, is currently not used in implementations of {\sc Max Cut} solvers. 
To the best of our knowledge, no research has been done on the efficiency of data reduction for {\sc Max Cut}, in particular with the goal of achieving small kernels \emph{in practice}. 

\medskip
\noindent
\textbf{Our Results.} We introduce new data reduction rules for the \textsc{Max Cut} problem, and show that nearly all previous reduction rules for the \textsc{Max Cut} problem can be encompassed by only four reduction rules.
Furthermore, we engineer efficient implementations of these reduction rules and show through extensive experiments we show that kernelization achieves a significant reduction on sparse graphs. 
Our experiments reveal that current state-of-the-art solvers can be sped up by up to \emph{multiple orders of magnitude} when combined with our data reduction rules. We achieve speedups on all instances tested. On social and biological networks in particular, kernelization enables us to solve four instances that were previously unsolved in a ten-hour time limit with state-of-the-art solvers; three of these instances are now solved in less than two seconds with our kernelization.

\section{Preliminaries}
\label{sec:Preliminaries}
Throughout this paper, we consider finite, simple and undirected graphs $G=(V,E)$ together with additive edge weight functions $\omega: E \to \MdR_{>0}$. 
For each vertex $v\in V$ let $N(v) := \{u\in V\mid \{v,u\}\in E\}$ denote its neighbors; its \emph{degree} in $G$ is $\mathsf{deg}(v):=|N(v)|$.
The {neighborhood} of a set $X \subseteq V$ is $N(X) := \bigcup_{v \in X} N(v) \setminus X$.
For a vertex set $S\subseteq V$, let $G[S]$ denote the subgraph of $G$ induced by $S$.
To specify the vertex and edge sets of a specific graph $G$, we use $V(G)$ and $E(G)$, respectively.
The set of edges between {the vertices of different} vertex sets $S_{1}, S_{2} \subseteq V$ is written as $E(S_{1},S_{2}) := E \cap (S_1 \times S_2)$.

For an integer $\ell$, a \emph{path} of length $\ell$ in $G$ is a sequence $P = \langle v_1,\ldots,v_{\ell+1}\rangle$ of distinct vertices such that $\{v_i,v_{i+1}\}\in E(G)$ for $i = 1,\ldots,\ell$.
A path with $v_{1}=v_{\ell}$ is called a \emph{cycle} of $G$.}
Graph $G$ is \emph{connected} {if there is a path from $v$ to $w$ for any pair $\{v,w\}$ of distinct vertices in $G$; and {\emph{disconnected}} otherwise. 
A \emph{connected component} of~$G$ is an inclusion-maximal connected subgraph of $G$.
For vertex sets $S \subseteq V(G)$, the set of \emph{external vertices} is {$C_{\textnormal{ext}}(S) = \{ v \in S \mid \exists w \in V(G) \setminus S, \{v,w\} \in E(G) \}$}, which is the set of vertices in $S$ that have some neighbor in $G$ outside $S$.
In similar fashion,~$C_{\textnormal{int}}(S) = S \setminus C_{\textnormal{ext}}(S)$ defines the set of \emph{internal vertices}. 

A \emph{clique} is a complete subgraph, and a \emph{near-clique} is a clique minus a single edge.
A \emph{clique tree} is a connected graph whose biconnected components are cliques, and a \emph{clique forest} is a graph whose connected components are clique trees. In such graphs, we use the term \emph{block} to refer to a biconnected component, bridge, or isolated vertex.
The class of \emph{clique-cycle forests} is defined as follows.
A clique is a clique-cycle forest, and so is a cycle.
The disjoint union of two clique-cycle forests is a clique-cycle forest.
In addition, a graph formed from a clique-cycle forest by identifying two vertices, each from a different (connected) component, is also a clique-cycle forest.

The {\sc Max Cut} problem is to find a vertex set $S\subseteq V$, such that $|E(S,V\setminus S)|$ is maximized. We denote the cardinality of a maximum cut by $\beta(G)$. At times, we may need to reason about a maximum cut given a fixed partitioning of a subset of $G$'s vertices. A partition of vertices $V'\subseteq V(G)$ is given as a $2$-coloring $\delta:V'\rightarrow \{0,1\}$. We let $\beta_\delta(G)$ denote the size of a maximum cut of $G$, given that $V'\subseteq V(G)$ is partitioned according to $\delta$.
{The {\sc Weighted Max Cut} problem is to find a vertex set $S$ of a given graph $G$ with additive weight function $\omega$ such that $\omega(E(S, V(G) \setminus S))$ is maximum}.
{The weight of a maximum cut is then given by $\beta(G,\omega) := \omega(E(S, V \setminus S))$.}
We denote {instances of the}~{\sc Max Cut} decision problem as {$(G,k)_{\textnormal{MC}}$}, where $G$ is a graph and $k \in \mathbb{N}_{0}$, 
If the size of a maximum cut in $G$ is $k$, then $(G,k)_{\textnormal{MC}}$ is a ``yes''-instance; otherwise, it is a ``no''-instance.

We address two more variations \textsc{Max Cut} in this paper.
The {\sc Vertex-Weighted Max Cut} problem takes as input a graph $G$ and two vertex weight functions $\omega_{0},\omega_1:V(G) \rightarrow \mathbb{R}$; the objective is to compute a bipartition~$V_{0} \cup V_{1} = V(G)$ that maximizes $|E(V_{0},V_{1})| + \sum_{v \in V_{0}} w_{0}(v) + \sum_{v \in V_{1}} w_{1}(v)$.
The \textsc{Signed Max Cut} problem takes as input a graph $G$ together with an edge labeling~$l:E(G) \rightarrow \{``\unaryplus",``\unaryminus"\}$; 
{the goal to find an $S \subseteq V(G)$ which maximizes the quantity~$\beta(G, l) := |E_{l}^{-}(S, V(G) \setminus S)| + |E^{+}(G[S], l) \cup E^{+}(G[V(G) \setminus S], l)|$, where~$E_{l}^{c}(S, V(G) \setminus S) := \{ e \in E(S, V(G) \setminus S) \mid l(e) = c \}$ and $E^{c}(l) := \{ e \in E(G) \mid l(e) = c \}$ for~$c \in \{``\unaryminus",``\unaryplus"\}$.
Similarly, for the neighborhood of a vertex (set), we use the notations~$N^{c}_{l}(v):=\{w \in V(G) \mid \{v,w\} \in E^{c}(l) \}$ and $N^{c}_{l}(X) := \bigcup_{v \in X} N^{c}_{l}(v) \setminus X$.
We call a triangle \emph{positive} if its number of ``\unaryminus''-edges is even.}
Any \textsc{Max Cut} instance can be {transformed into} a \textsc{Signed Max Cut} instance by labeling all edges with ``\unaryminus''.

Let $\Sigma^{*}$ denote the set of input instances for a decision problem.
A parameterized problem $\Pi \subseteq \Sigma^{*} \times \mathbb{N}$ is \emph{fixed-parameter tractable} if there is an algorithm~$\mathcal{A}$ (called a \emph{fixed-parameter algorithm}) that decides membership in $\Pi$ for any input pair $(x, k) \in \Sigma^*\times\mathbb N$ in time~$f(k) \cdot |(x,k)|^{O(1)}$ for some computable function $f : \mathbb{N} \rightarrow \mathbb{N}$. 

{A \emph{data reduction rule} (often shortened to \emph{reduction rule}) for a parameterized problem $\Pi$ is a function $\phi : \Sigma^{*} \times \mathbb{N} \rightarrow \Sigma^{*} \times \mathbb{N}$ that maps an instance $(x, k)$ of $\Pi$ to an equivalent instance $(x',k')$ of $\Pi$ such that $\phi$ is computable in time polynomial in $|x|$ and~$k$.
We call two instances of $\Pi$ \emph{equivalent} if either both or none belong to $\Pi$.
Observe that for two equivalent ``yes''-instances $(G,\beta(G))$ and $(G',\beta(G'))$, the relationship~$\beta(G) = \beta(G') + k$ holds for some $k \in \mathbb{Z}$. 

\subsection{Related Work}
\label{sec:PreviousResearch}
{Several studies have been made } in the direction of providing fixed-parameter algorithms for the {\sc Max Cut} problem {\cite{DBLP:journals/corr/abs-1212-6848,DBLP:journals/corr/abs-1112-3506,Etscheid2018,DBLP:conf/csr/Madathil0Z18}}.
Among these, a fair amount of kernelization rules have been introduced with the goal of effectively reducing {\sc Max Cut} instances \cite{DBLP:journals/corr/abs-1212-6848,DBLP:journals/corr/abs-1112-3506,Etscheid2018,DBLP:conf/csr/Madathil0Z18,Prieto:2005:MES:1082260.1082274,DBLP:journals/corr/FariaKSS15}.
Those reductions typically have some constraints on the subgraphs, like being clique forests or clique-cycle forest.
Later, we propose a new set of reductions that does not need this property and cover most of the known reductions ~\cite{DBLP:journals/corr/abs-1112-3506,Etscheid2018,DBLP:conf/csr/Madathil0Z18,DBLP:journals/corr/FariaKSS15}.
There are other reductions rules that are fairly simplistic and focus on very narrow cases~\cite{Prieto:2005:MES:1082260.1082274}. 
We now explain the Edwards-Erd{\H{o}}s bound and the spanning tree bound.

\textbf{Edwards-Erd{\H{o}}s Bound.}
\label{ee:aee}
{For a connected graph}, the Edwards-Erd{\H{o}}s bound \cite{edwards1973some,edwards1975improved} is defined as $EE(G) = \frac{|E(G)|}{2} + \frac{|V(G)| - 1}{4}$. 
A linear-time {algorithm that computes a cut satisfying the Edwards-Erd{\H{o}}s bound for} any given graph is provided by Van Ngoc and Tuza~\cite{van1993linear}. 
The \textsc{Max Cut Above Edwards-Erd{\H{o}}s} (\textsc{Max Cut AEE}) problem asks for a graph $G$ and integer $k \in \mathbb{N}_{0}$ if $G$ admits a cut of size~$EE(G) + k$. 
All kernelization rules for \textsc{Max Cut AEE} require  a set $S \subseteq V$ set such that $G - S$ is a clique forest.  
Etscheid and Mnich~\cite{Etscheid2018} propose an algorithm that computes such a set $S$ of at most $3k$ vertices in time $O(k \cdot(|V| + |E|))$.

\textbf{Spanning Tree Bound.} \label{st:aee} Another approach is based on utilizing the spanning forest of a graph \cite{DBLP:conf/csr/Madathil0Z18}.
For a given $k \in \mathbb{N}_{0}$, a {\sc Max Cut} of size $|V| - 1 + k$ is searched for.
This decision problem is denoted as \textsc{Max Cut AST} (\textsc{Max Cut Above Spanning Tree}).
For sparse graphs, this bound is larger than the Edwards-Erd{\H{o}}s bound. 
The reductions for the problem require a set $S \subset V(G)$ such that $G - S$ is a clique-cycle forest.

\section{New Data Reduction Rules}
We now introduce our new data reduction rules and prove their correctness.
The main feature of our new rules is that they do not depend on the computation of a clique-forest to determine if they can be applied.
Furthermore, our new rules subsume almost all rules from previous works~\cite{DBLP:journals/corr/abs-1212-6848,DBLP:journals/corr/abs-1112-3506,Etscheid2018,DBLP:conf/csr/Madathil0Z18,DBLP:journals/corr/FariaKSS15}
with the exception of Reduction Rules~10 and~11 by Crowston et al.~\cite{DBLP:journals/corr/abs-1212-6848}. 
We provide details in 
\cite{maFerizovic}.
For an overview of how rules are subsumed, consult Table~\ref{tab:inclusions}.
Hence, our algorithm will only apply the rules proposed in this section.
We provide proofs for the rules that proved most useful in our experimental
evaluation. 

\begin{table}[b]
  \centering
  \caption{Reduction rules from previous work subsumed by our new rules.
  A \checkmark~in row $a$ and column~$b$ means that the rule from row $a$ subsumes the rule from column $b$.
  If there are multiple \checkmark{}s in a column (say, rows $a$ and $b$ in column $c$), then rules $a$ and $b$ combined subsume rule $c$.\label{tab:inclusions}}
  \setlength{\tabcolsep}{.7ex}
  \begin{tabular}{l@{\hskip 18pt}c@{\hskip 18pt}ccc@{\hskip 18pt}cc@{\hskip 18pt}c@{\hskip 18pt}cccccccc}
    \toprule
    Source             & \cite{DBLP:journals/corr/FariaKSS15} & \multicolumn{3}{c}{\hspace*{-15pt}\cite{DBLP:journals/corr/abs-1112-3506}} & \multicolumn{2}{c}{\hspace*{-15pt}\cite{DBLP:journals/corr/abs-1212-6848}} & \cite{Etscheid2018} & \multicolumn{8}{c}{\hspace*{-11pt}\cite{DBLP:conf/csr/Madathil0Z18}}                                                 \\
    \midrule
    Rule           & A                                    & 5                  & 6                  & 7                 & 8 & 9                                       & 9                   & 6          & 7          & 8          & 9          & 10         & 11         & 12         & 13         \\
    \midrule
    \ref{rule:8+}      &                                      &                    & \checkmark         &                   & \checkmark &                                          &                     &            &            &            &            &            &            &            &            \\
    \ref{rule:x1}      & \checkmark                           & \checkmark         &                    & \checkmark        & & \checkmark                              & \checkmark          & \checkmark & \checkmark & \checkmark & \checkmark &            &            & \checkmark & \checkmark \\
    \ref{rule:s5}      &                                      &                    &                    &                   & &                                        &                     & \checkmark &            & \checkmark & \checkmark & \checkmark & \checkmark &            &            \\
    \ref{r:rem:shared} &                                      &                    & \checkmark         &                   &  &                                       &                     &            &            &            &            &            &            &            &           \\
    \bottomrule
  \end{tabular}
\end{table}

\begin{KernelizationRule}
\label{rule:x1}
  Let $G=(V,E)$ be a graph and let $S \subseteq V$ induce a clique in $G$.
  If~$|C_{\textnormal{ext}(G)}(S)| \leq \ceil*{|S|/2}$, then $\beta(G) = \beta(G') + \beta(K_{|S|})$ for $G'=(V \setminus C_{\textnormal{int}(G)}(S),E \setminus E(G[S]))$.\hfill{}
\end{KernelizationRule}
\begin{proof}
Note that any partition of the clique $G[S]$ into two vertex sets of size $\lceil |S|/2\rceil$ and $\lfloor |S|/2\rfloor$ is a maximum cut of $G[S]$.
Suppose we fix the partitions of the at most $\lceil |S|/2 \rceil$ external vertices of $S$.
Then the at least $\lfloor |S|/2\rfloor$ internal vertices can be assigned to the partitions so they each contain $\lceil |S|/2 \rceil$ and $\lfloor |S|/2\rfloor$ vertices.
  Thus, regardless of how $C_{\textnormal{ext}(G)}(S)$ is partitioned, the size of a maximum cut of $G[S]$ remains the same. 
\end{proof}

We can exhaustively apply Reduction Rule~\ref{rule:x1} in~$O(|V| \cdot \Delta^{2})$ time by scanning over all vertices in the graph.
When scanning vertex $v$, we check whether $N(v) \cup \{v\}$ induces a clique.
This finds all cliques with at least one internal vertex.
Checking whether Reduction Rule~\ref{rule:x1} is applicable is then straightforward by counting the number of vertices with degree higher than the size of the clique.

\begin{KernelizationRule}
\label{rule:s5}
  Let $(a',a,b,b')$ be an induced $3$-path in a graph $G$ with $N(a) = \{a',b\}$ and $N(b) = \{a,b'\}$.
  Construct $G'$ from $G$ by adding a new edge $\{a',b'\}$ and removing the vertices $a$ and $b$.
  Then $\beta(G) = \beta(G') + 2$.
\end{KernelizationRule}
\begin{proof}
  Let $S = \{a',a,b,b'\}$ and let $\delta:V\rightarrow\{0,1\}$ be an assignment of vertices to the partitions of a cut in $G$.
  We distinguish two cases:
  \begin{itemize}
    \item{Case $\delta(a') = \delta(b')$:} If $\delta(a) = \delta(b) = \delta(a')$, then no edges of $G[S]$ are cut. Notice that this cut is not maximum since moving $b$ between partitions increases the cut size by two.
      If $\delta(a) \neq \delta(b)$, then exactly two edges in $G[S]$ are cut. 
    \item{Case $\delta(a') \neq \delta(b')$:} By choosing $\delta(a) = \delta(b')$ and $\delta(b) = \delta(a')$, all three edges in $G[S]$ are cut.
      In $G'$, the edge between $a'$ and $b'$ is cut, so $\beta(G) = \beta(G') + 2$. \qedhere%
  \end{itemize}%
\end{proof}

\begin{KernelizationRule}
\label{rule:add:cliques}
  Let $G$ be a graph and let $S\subseteq V(G)$ induce a near-clique in~$G$.
  Let $G'$ be the graph obtained from $G$ by adding the missing edge $e'$ so that $S$ induces a clique in $G'$.
  If $|S|$ is odd or $|C_{\textnormal{int}(G)}(S)| > 2$, then~$\beta(G) = \beta(G')$.
\end{KernelizationRule}
\begin{proof}
  Let $(u, v)$ be the edge added to the graph and $\delta$ any 2-coloring of $C_{\textnormal{ext}(G)}(S)$.
  {We show that a maximum cut of~$G'$ exists such that $u$ and $v$ are in the same partition.
  As $G$ has one less edge than $G'$, this means that~$\beta_{\delta}(G[S]) = \beta_{\delta}(G'[S])$, which implies
  that $\beta(G) = \beta(G')$.}
	
  Define $V_{c} = \{x \in C_{\textnormal{ext}(G')}(S) \mid \delta(x) = c \}$ for $c \in \{0, 1\}$.
  Without loss of generality, assume $|V_0| \leq |V_1|$.
  Note that, given the partition for $C_{\textnormal{ext}(G')}(S)$, maximizing
  the cut of $S$ means minimizing $||V_0| - |V_1||$. We distinguish three cases:
  \begin{itemize}
  \item{$|V_0| - |V_1| \leq 2$:} By adding $u$ and $v$ to $V_0$, $||V_0| -
    |V_1||$ decreases. The rest of the internal vertices have to be distributed
    among $V_0$ and $V_1$ such that $||V_0| - |V_1||$ is minimized 
  \item{$|V_0| - |V_1| = 1$:} By adding $u$ and $v$ to $V_0$, $||V_0| -
    |V_1||$ stays $1$. If $|S|$ is odd, then $1$ is the minimal value possible and
    $|C_{\textnormal{int}(G)}(S)|$ is even. So the remaining internal vertices can be
    distributed evenly between $V_0$ and $V_1$. If $S$ is even, then an odd
    number of internal vertices are left (and at least one by
    the definition of the rule) which can be distributed to balance $V_0$ and $V_1$.
  \item{$|V_0| = |V_1|$:} By adding $u$ and $v$ to $V_0$, $||V_0| -
    |V_1||$ becomes 2. If $|S|$ is odd, then an odd number of internal vertices
    is left to assign to such that $||V_0| - |V_1||$ becomes 1. If $|S|$ is even
    then there is an even number of internal vertices left which can be distributed to balance $V_0$ and $V_1$.\qedhere
  \end{itemize}
\end{proof}

Since some cliques are irreducible by currently known rules, it may be beneficial to also apply Reduction Rule~\ref{rule:add:cliques} `in reverse'.
Although this `reverse' reduction neither reduces the vertex set nor (as our experiments suggest) lead to applications of other rules,
it can undo unfruitful additions of edges made by Reduction Rule~\ref{rule:add:cliques} and may remove other edges from the graph.

\begin{KernelizationRule}
\label{rule:rem:cliques}
  Let $G$ be a graph and let $S \subseteq V(G)$ induce a clique in $G$.
  If $|S|$ is odd or $C_{\textnormal{int}(G)}(S) > 2$, an edge between two vertices of $C_{\textnormal{int}(G)}(S)$ is removable.
  That is, $\beta(G) = \beta(G')$ for $G'=(V,E \setminus \{e\})$, $e \in E(G[C_{\textnormal{int}(G)}(S)])$. 
\end{KernelizationRule}
\begin{proof}
  Follows from the correctness of Reduction Rule~\ref{rule:add:cliques}.
\end{proof}

The following reduction rule is closely related to the upcoming generalization of Reduction Rule 8 by Crowston et al.~\cite{DBLP:journals/corr/abs-1212-6848}.
It is able to further reduce the case where $|X| = |N(X)|$ for a clique $X$ of~$G$.
In comparison, the generalization of Reduction Rule 8 from \cite{DBLP:journals/corr/abs-1212-6848} is able to handle the case $|X| > |N(X)|$.
Due to the degree by which these rules are similar, they are also merged together in our implementation, as the techniques to handle  both are the same.

\begin{KernelizationRule}\label{r:rem:shared}
  Let $X\subseteq V$ induce a clique in a graph $G$, where~$|X| = |N(X)| \geq 1$ and $N(X) =N(x)\setminus X $ for all $x \in X$.
  Create $G'$ from~$G$ by removing an arbitrary vertex of $X$.
  Then $\beta(G) = \beta(G') + |X|$.
\end{KernelizationRule}
\begin{proof}
  Let $S := X \cup N_{G}(X)$ and $\delta$ be any 2-coloring of $N_{G}(X)$.
  Note that $C_{\textnormal{ext}(G)}(S) \subseteq N_{G}(X)$ -- the removal of $N_{G}(X)$ disconnects $X$ from the remainder of the graph.
	
  Define $V_{c} = \{x \in N_G(X) \mid \delta(x) = c \}$ and $z_{c} := |V_{c}|$ for $c \in \{0, 1\}$.
  We distribute the vertices in $X$ among $V_0$ and $V_1$ such that
  $E(V_{0},V_{1})$ is maximized.
  Notice that every vertex in $X$ is connected to all other vertices in~$S$.
  The size of any cut is therefore~$p(c_{0},c_{1}) = c_{0} z_{1} + c_{1} z_{0} + c_{0} c_{1} + |E(V_{0},V_{1})|$, where $c_{0}$ and $c_{1}$ denote the number of vertices from $X$ that we want to insert into~$V_{0}$ and $V_{1}$, respectively.
  This can be rewritten as $p(c_{0},c_{1}) = (z_{0} + c_{0}) \cdot (z_{1} + c_{1}) - z_{0} z_{1} + |E(V_{0},V_{1})|$.
  As all other parts are constant, this reduces to maximizing $(z_{0} + c_{0})
  \cdot (z_{1} + c_{1})$. As $ z_{0} + c_{0}  +  z_{1} + c_{1} $ is constant, $(z_{0} + c_{0})
  \cdot (z_{1} + c_{1})$ is maximized when $|(z_{0} + c_{0}) - (z_{1} + c_{1})|$
  is minimized. 

  Because $|X| = |N_{G}(X)|$, it is always possible to distribute the vertices
  of $X$ such that $z_{0} + c_{0} = z_{1} + c_{1} = |X|$, which then maximizes $p(c_{0},c_{1})$.
  Removing any vertex $x \in X$ from $G$ will change the cut by~$-|X|$:
  without loss of generality, let $x \in V_{0}$. Then $|X| + |N_{G}(X)|$ is odd and $|z_{0} +
  (c_{0}-1) - z_{1} + c_{1}| = 1$, which maximizes the cut.
  Then,~$p(c_{0} - 1, c_{1}) = p(c_{0},c_{1}) - |X|$.
\end{proof}

The following algorithm identifies all candidates of Reduction Rule~\ref{r:rem:shared} in linear time. 
First, we order the adjacencies of all vertices.
That is, for every vertex $v\in V$, the vertices in $N(v)$ are sorted according to a numeric identifier assigned to every vertex. For this, we create an auxiliary array of empty lists of size $|V(G)|$.
We then traverse the vertices $w \in N(v)$ for every vertex $v \in V(G)$ and insert each  pair $(v,w)$ in a list identified by indexing the auxiliary array with $w$. We then iterate once over the array from the lowest identifier to the highest and recreate the graph with sorted adjacencies.
In total, this process takes $O(|V| + |E|)$ time.

For any clique $X$ of $G$, we have to check if for all pairs $(x_{1}, x_{2})$ of vertices from $X$ that
$N(x_{1}) \cup \{x_{1}\}= N(x_{2}) \cup \{x_{2}\}$ holds (neighborhood condition).
Our algorithm uses tries \cite{fredkin1960trie,de1959file} to find all candidates.
A trie supports two operations, \textsc{Insert(key,val)} and \textsc{Retrieve(key)}.
The \textsc{key} parameter is an array of integers and \textsc{val} is a single integer.
Function \textsc{Retrieve} returns all inserted values by \textsc{Insert} that have the same key.
Internally, a trie stores the inserted elements as a tree, where every node corresponds to one integer of the key and every prefix is stored only once.
That means that two keys sharing a prefix share the same path through the trie until the position where they differ.

For each vertex $v \in V$, we use the ordered set $N(v) \cup \{v\}$ as \textsc{key} and $v$ as the \textsc{val} parameter. Notice that $N(v)$ is already sorted. The key $N(v) \cup \{v\}$ can be then computed through an insertion of $v$ into the sequence $N(v)$ in time $O(|N(v)|)$.
After \textsc{Insert($N(v) \cup \{v\}$,$v$)} is done for every vertex $v\in V$, each trie leaf contains all vertices that satisfy the condition of Reduction Rule~\ref{r:rem:shared}.
Meaning, for every vertex pair $(x_{1},x_{2})$ of a trie leaf, the neighborhood condition is met.
We then verify whether the vertex set $X$ of a leaf is a clique, in~$O(|E(X)|)$ time.
As each such set~$X$ is considered exactly once and the graph is fully partitioned, this requires~$O(|V| + |E|)$ time in total.
As a last step, we check whether $|X| > \max\{|N(X)|, 1\}$ by using the observation that $\forall x \in X: |N(X)| = \mathsf{deg}(x) - |X|$.
{In Sect.~\ref{sec:Implementation}, we describe a timestamping system that assists the above procedure in not having to repeatedly check the same structures after any amount of vertices and edges are added or removed from $G$.
However, in those later applicability checks, we disregard sorting the adjacencies of all vertices in linear time again.
Rather we simply use a comparison based sort on the adjacencies.}

\medskip
The next reduction rule is our only rule whose application turns unweighted instances into instances of \textsc{Weighted Max Cut}.
Our experiments show that this can reduce the kernel size significantly.
This is noteworthy, given that existing solvers for {\sc Max Cut} usually support weighted instances.

\begin{KernelizationRule}
\label{r:weighted:compression}
  Let $G$ be a graph, $w : E \rightarrow \mathbb{Z}$ a weight function, and
  $(a,b,a')$ be an induced 2-path with $N(b)=\{a,a'\}$.
  Let $e_{1}$ be the edge between vertex $a$ and $b$; let~$e_{2}$ be the one between $b$ and $a'$.
  Construct $G'$ from $G$ by deleting vertex $b$ and adding a new edge~$\{a,a'\}$ with $w'(\{a,a'\})= \max \{w(e_{1}), w(e_{2})\} - \max \{0, w(e_{1})+w(e_{2})\}$.
  Then $\beta(G,w) = \beta(G',w) + \max \{0, w(e_{1})+w(e_{2})\}$.
\end{KernelizationRule}
\begin{proof}
  Let $\delta$ be a maximum cut of $G$ and consider the following two cases:
  \begin{itemize}
	\item{$\delta(a) = \delta(a')$:} If $w(e_{1})+w(e_{2}) > 0$, then
      $\delta(b) \neq \delta(a)$. Otherwise, $\delta(b) = \delta(a)$. In total,
      the path contributes $\max \{0, w(e_{1})+w(e_{2})\}$ to the cut. in $G'$,
      the edge between $a$ and $a'$ is not cut, so $\beta(G,w) = \beta(G',w') + \max \{0, w(e_{1})+w(e_{2})\}$.
    \item{$\delta(a) \neq \delta(a')$:} If $w(e_1) > w(e_2)$, then $\delta(b) =
      \delta(a')$. Otherwise, $\delta(b) = \delta(a)$. In total, the path
      contributes $\max \{w(e_{1}), w(e_{2})\}$ to the cut. In $G'$, the edge
      between $a$ and $a'$ is cut and contributes $w'(\{a,a'\})= \max
      \{w(e_{1}), w(e_{2})\} - \max \{0, w(e_{1})+w(e_{2})\}$ to the cut, so
      again $\beta(G,w) = \beta(G',w') + \max \{0, w(e_{1})+w(e_{2})\}$.\qedhere
  \end{itemize}
\end{proof}

Our next two rules (Reduction Rules~\ref{rule:8+} and~\ref{rule:8s+}) generalize Reduction Rule 8 by Crowston et al.~\cite{DBLP:journals/corr/abs-1212-6848}, which we restate for completeness. 

\addtocounter{KernelizationRule}{1}
\begin{KernelizationRule}[(\cite{DBLP:journals/corr/abs-1212-6848}, Reduction Rule~8)]
\label{rule:8}
	
  Let $(G,l)$ be a signed graph, $S\subseteq V$ a set of vertices such that $G[V\setminus S]$ is a clique forest, and $C$ a block in $G[V \setminus S]$.
  If there is a~$X \subseteq C_{\textnormal{int}(G[V\setminus S])}(C)$ such that $|X| > \frac{|C|+|N(X)\cap S|}{2} \geq 1$, {$N_{l}^{+}(x) \cap S = N_{l}^{+}(X) \cap S$ and~$N_{l}^{-}(x) \cap S = N_{l}^{-}(X) \cap S$} for all $x \in X$. Construct the graph $G'$ from $G$ by removing any two vertices $x_{1}, x_{2} \in X$, then $\beta(G') - EE(G') = \beta(G) - EE(G)$.
\end{KernelizationRule}

Note that, for unsigned graphs, $N_{l}^{+}(x) = \emptyset$ and~$N_{l}^{-}(x) = N(x)$ for every vertex $x$.

Here, different choices of~$S$ lead to different applications of this rule.
Our generalizations do not require such a set anymore and can find \emph{all} possible applications for any choice of $S$.

\addtocounter{KernelizationRule}{-2}
\newcommand{\trEightPlus}{
Let $X$ be the vertex set of a clique in $G$ with $|X| > \max\{|N(X)|,1\}$ and $N(X) =N(x)\setminus X $ for all $x \in X$. Construct the graph $G'$ by deleting two arbitrary vertices~$x_{1}, x_{2} \in X$ from $G$. Then~$\beta(G) = \beta(G') + |N(x_{1})|$.
}

\begin{ExactKernelizationRule}{\ref{rule:8s+}${_{w=1}}$}\label{rule:8+}
\trEightPlus
\end{ExactKernelizationRule}

We show the correctness of Reduction Rule~\ref{rule:8+} by reducing it 
to Reduction Rule 8 by Crowston et al.~\cite{DBLP:journals/corr/abs-1212-6848}.

\begin{proof}
Let $S = V\setminus X$ and $C = X$. Since $X$ is a clique, $G[V\setminus S]$ is a clique forest. From $|X| > \max\{|N(X)|,1\}$
it follows that $|X| > \frac{|X|+|N(X)|}{2} = \frac{|C|+|N(X) \cap S|}{2} \geq 1$. Also, $N(x) \setminus X =
N(x) \cap S$ and $N(X) \cap S = N(X)$, so all conditions for Reduction Rule~\ref{rule:8} are satisfied.

It remains to show that $\beta(G) = \beta(G') + |N(x_1)|$.
Note that $|E(G')| = |E(G)| - |N_{G}(x_{1})| - (|N_{G}(x_{2})| - 1)$ and $|V(G')| = |V(G)| - 2$. By Reduction Rule~\ref{rule:8}, we know that $\beta(G') - EE(G') = \beta(G) - EE(G)$, therefore we have that

{\small
\begin{align}
  \beta(G) - \beta(G') 
  &= EE(G) - EE(G') \nonumber\\
  &= \frac{|E(G)|}{2} + \frac{|V(G)| - 1}{4} - \left(\frac{|E(G')|}{2} + \frac{|V(G')| - 1}{4}\right)\nonumber\\
  &=\frac{|E(G)|}{2} + \frac{|V(G)| - 1}{4} - \left( \frac{|E(G)| - |N_{G}(x_{1})| - (|N_{G}(x_{2})| - 1) }{2} - \frac{(|V(G)| - 2) - 1}{4} \right)\nonumber\\
  &= \frac{(|V(G)| - 1) - |V(G)| +2 +1}{4} - \frac{- |N_{G}(x_{1})| - (|N_{G}(x_{2})| - 1) }{2}\nonumber\\
  &=
  \label{eq:nx1nx2}
\frac{2}{4} - \frac{- |N_{G}(x_{1})| - |N_{G}(x_{1})| + 1 }{2} \\
  &= \frac{2}{4} - \frac{- 2|N_{G}(x_{1})| + 1 }{2} \nonumber\\
  &= \frac{2}{4} - \frac{1}{2} + |N_{G}(x_{1})| \nonumber\\
  &= |N_{G}(x_{1})|.\nonumber
\end{align}
}
Where~\eqref{eq:nx1nx2} follows from $N_{G}(x_{1}) = N_{G}(x_{2})$.
\end{proof}

\begin{KernelizationRule}
\label{rule:8s+}
  Let $X\subseteq V$ induce a clique in a signed graph $(G,l)$ such
  that~$\forall e \in E(X) : l(e) = ``\unaryminus"$ and  $|X| >
  \max\{|N(X)|,1\}$, $N_{l}^{+}(X) =N_{l}^{+}(x)\setminus X $, and~$N_{l}^{-}(X)
  =N_{l}^{-}(x)\setminus X $ for all $x \in X$. 
  Construct $G'$ by deleting two arbitrary vertices~$x_{1}, x_{2} \in X$ from $G$.
  Then $\beta(G) = \beta(G') + |N(x_{1})|$.
\end{KernelizationRule}
\begin{proof}[Proof (Sketch).]
  The proof for this rule is almost identical to the proof of Reduction Rule~\ref{rule:8+}.
\end{proof}

Using an almost equivalent approach as we did for Reduction Rule~\ref{r:rem:shared}, we can find all candidates of this reduction rule in linear time.

In order to also reduce weighted instances to some degree, we use a simple weighted scaling of two reduction rules.
That is, we extend their applicability from an unweighted subgraph to a subgraph where all edges have the same weight $c \in \mathbb{R}$.
We do this for Reduction Rules~\ref{rule:x1} and~\ref{rule:add:cliques}.

\begin{ExactKernelizationRule}{\ref{rule:x1}$_{w=c}$}
\label{rule:x1:scaled}
  Let $(G,\omega)$ be a weighted graph and let $S \subseteq V(G)$ induce a clique with $\omega(e)=c$ for every edge $e \in E(G[S])$ for some constant $c \in \mathbb{R}$.
  Let $G'=(V(G) \setminus C_{\textnormal{int}(G)}(S),E(G) \setminus E(G[S]))$ with $\omega'(e) = \omega(e)$ for every $e \in E(G')$.
  If~$|C_{\textnormal{ext}(G)}(S)| \leq \ceil*{\frac{|S|}{2}}$, then $\beta(G,\omega) = \beta(G',\omega') + c \cdot \beta(K_{|S|})$.
\end{ExactKernelizationRule}

\begin{ExactKernelizationRule}{\ref{rule:add:cliques}$_{w=c}$}
\label{rule:add:cliques:scaled}
  Let $(G,\omega)$ be a weighted graph and let $S\subseteq V(G)$ induce a near-clique in $G$.
  Furthermore, let $\omega(e)=c$ for every edge~$e \in E(G[S])$ for some constant $c \in \mathbb{R}$.
  Let $G'$ be the graph obtained from $G$ by adding the edge $e'$ so that $S$ induces a clique in $G'$.
  Set $\omega'(e')=c$, and~$\omega'(e)=\omega(e)$ for~$e \in E(G)$.
  If $|S|$ is odd or $|C_{\textnormal{int}(G)}(S)| > 2$, then~$\beta(G,\omega) = \beta(G',\omega')$.
\end{ExactKernelizationRule}

\section{Implementation}
\label{sec:Implementation}

\subsection{Kernelization Framework}
We now discuss our \emph{overall} kernelization framework in detail.
Our algorithm begins by generating an unweighted instance by replacing every weighted edge by an unweighted subgraph with a specific structure.
Afterwards, we apply our full set of unweighted reduction rules: \ref{rule:x1}, \ref{rule:8+} (together with \ref{r:rem:shared}), \ref{rule:s5}, and \ref{rule:add:cliques}.
As already mentioned earlier, Reduction Rule~\ref{rule:8+} is the unweighted version of \ref{rule:8s+}.
We then create a signed instance of the graph by exhaustively executing weighted path compression using Reduction Rule~\ref{r:weighted:compression} with the restriction that the resulting weights are $-1$ or $+1$.
We then exhaustively apply Reduction Rule~\ref{rule:8s+}. 
Once the signed reductions are done, we apply Reduction Rule~\ref{r:weighted:compression} to fully compress all paths into weighted edges. 
This is then succeeded by Reduction Rule~\ref{rule:x1:scaled} and \ref{rule:add:cliques:scaled}.
We then transform the instance into an unweighted one and apply Reduction Rule~\ref{rule:rem:cliques} in order to avoid cyclic interactions between itself and Reduction Rule~\ref{rule:add:cliques}.
Finally, if a weighted solver is to be used on the kernel, we exhaustively perform Reduction Rule~\ref{r:weighted:compression} to produce a weighted kernel.
Note that different permutations of the order in which reduction rules are applied can lead to different results. 

\subsection{Timestamping}
Next we describe how to avoid unnecessary checks for the applicability of reduction rules.
For this purpose, let the time of the most recent change in the neighborhood of a vertex be $T : V(G) \rightarrow \mathbb{N}_{0}$ and let the variable $t \in \mathbb{N}$ describe the current time.
Initially, $T(v) = 0, \forall v \in V$ and $t = 1$. 
Every time a reduction rule performs a change on $N(v)$, set~$T(v) = t$ and increment $t$.
For each individual Reduction Rule $r$, we also maintain a timestamp $t_{r} \in \mathbb{N}_{0}$ (initialized with $0$), indicating the upper bound up to which all vertices have already been processes.
Hence, all vertices $v \in V$ with $T(v) \leq t_{r}$ do not need to be checked again by Reduction Rule~$r$. 
Note that timestamping only works for ``local'' reduction rules---the rules whose applicability can be determined by investigating the neighborhood of a vertex.
Therefore, we only use this technique for Reduction Rules~\ref{rule:x1} and \ref{rule:8s+}. 

\section{Experimental Evaluation}
\label{sec:Evaluation}
\subsection{Methodology and Setup}
All of our experiments were run on a machine with four Octa-Core Intel Xeon E5-4640 processors running at 2.40GHz CPUs with $512$ GB of main memory. 
The machine runs Ubuntu 18.04.
All algorithms were implemented in C++ and compiled using gcc version~7.3.0 with optimization flag \texttt{-O3}. 
We use the following state-of-the-art   \textsc{Weighted Max Cut} solvers for comparisons: the exact solvers \textsc{LocalSolver}~\cite{localsolverx1} (heuristically finds a large cut, and can then verify if it is maximum), Biq Mac~\cite{RRW10} as well as the heuristic solver \textsc{MqLib}~\cite{dunning-2018}.
\textsc{MqLib} is unable to determine on its own when it reaches a maximum cut and always exhausts the given time limit.
We also evaluated an implementation of the reduction rules used by Etscheid and Mnich~\cite{Etscheid2018}; however, preliminary experiments indicated that it performs worse than current state-of-the-art solvers.
In the following, for a graph $G=(V,E)$, $G_{\textnormal{ker}}$ denotes the graph after all reductions have been applied exhaustively.
For this purpose, we examine the following efficiency metric: we denote the \emph{kernelization efficiency} by {$e(G) = 1 - |V(G_{\textnormal{ker}})|/{|V(G)|}$}. Note that $e(G)$ is $1$ when all vertices are removed after applying all reduction rules, and $0$ if no vertices are removed.

\label{datasets}
For our experiments we use four different datasets:
First, we use random instances from four different graph models that were generated using the KaGen graph generator~\cite{funke2019communication,sanders2016generators}. 
In particular, we used \erdos~graphs (GNM), random geometric graphs (RGG2D), random hyperbolic graphs (RHG) and Barab\'{a}si-Albert graphs (BA).
The main purpose of these instances is to study the effectiveness of individual reduction rules for a variety of graph densities and degree distributions.
To analyze the practical impact of our algorithm on current-state-of-the-art solvers we use a selection of sparse real-world instances by Rossi and Ahmed~\mycite{nr}, as well as instances from VLSI design (\texttt{g00*}) and image segmentation (\texttt{imgseg-*}) by Dunning et al.~\mycite{dunning-2018}.
Note that the original instances by Dunning et al.~\mycite{dunning-2018} use floating-point weights that we scaled to integer weights.
Finally, we evaluate denser instances taken from the rudy category of the \textsc{Biq Mac} Library~\cite{biqmac}.
We further subdivide these instances into medium- and large-sized instances.

\subsection{Performance of Individual Rules}
To analyze the impact of each individual reduction rule, we measure the size of the kernel our algorithm procedures before and after their removal. 
Fig.~\ref{kernelization:cmp:iso:gnm} shows our results on RGG2D and GNM graphs with $2048$ vertices and varying density.
We have settled on those two types of graphs as they represent different ends on the spectrum of kernelization efficiency.
In particular, kernelization performs good on instances that are sparse and have a non-uniform degree distribution.
Such properties are given by the random geometric graph model used for generating the RGG2D instances.
Likewise, kernelization performs poor on the uniform random graphs that make up the GNM instances.
We excluded Reduction Rule~\ref{rule:rem:cliques} from these experiments as it only removes edges and thus leads to now difference in the kernelization efficiency. 

Looking at Fig.~\ref{kernelization:cmp:iso:gnm}, we can see that Reduction Rule~\ref{rule:x1} gives the most significant reduction in size. 
Its absence always diminishes the result more than any other rule. 
In particular, we see a difference in efficiency of up to $0.47$ (RGG2D) and $0.41$ (GNM) when removing Reduction Rule~\ref{rule:x1}.
The second most impactful rule for the RGG2D instances is Reduction Rule~\ref{rule:8s+} with a difference of only up to $0.04$.
For the GNM instances Reduction Rule~\ref{rule:s5} is second with a difference of up to $0.17$.
However, note that Reduction Rules~\ref{rule:add:cliques} and \ref{rule:8s+} lead to no difference in efficiency on these instances.
Thus, we can conclude that depending on the graph type, different reduction rules have varying importance.
Furthermore, our simple Reduction Rule~\ref{rule:x1} seems to have the most significant impact on the overall kernelization efficiency. 
Note that this is in line with the theoretical results from Table~\ref{tab:inclusions}, which states that Reduction Rule~\ref{rule:x1} covers most of the previously published reduction rules and Reduction Rule~\ref{rule:s5} still covers many but less rules from previous work.

\begin{figure}[t]
  \centering
  \includegraphics[width=1.0\textwidth]{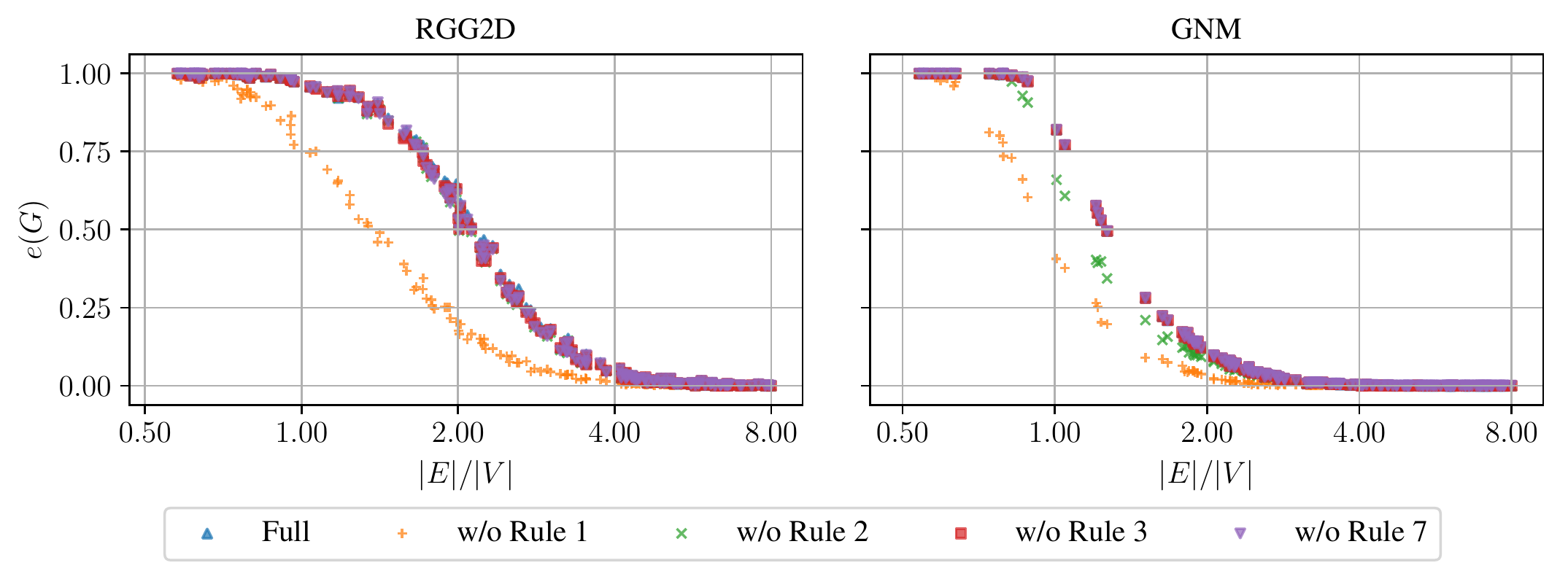}
  \caption{Tests consist of 150 synthetic instances. We compare the kernelization efficiency of our full algorithm to the efficiency of our algorithm without a particular reduction rule.\label{kernelization:cmp:iso:gnm}}
\end{figure}

\subsection{Exactly Computing a Maximum Cut}
\label{subsec:CutEvaluation}
To examine the improvements kernelization brings for medium-sized instances, we compare the time required to obtain a maximum cut for both the kernelized and the original instance.
We performed these experiments using both \textsc{LocalSolver} and \textsc{Biq Mac}.
Note that we did not use \textsc{MqLib} as it is not able to verify the optimality of the cut it computes.
The results of our experiments for our set of real-world instances are given in Table~\ref{solver:weighted} (with weighted path compression) and Table~\ref{solver:unweighted} (without weighted path compression).
Since the image segmentation instances are already weighted, they are omitted from Table~\ref{solver:unweighted}.
It is noteworthy that we do not include the results for the rudy instances from the Biq Mac library.
These instances feature a uniform edge distribution and an overall average degree of at least $3.5$.
Our preliminary experiments indicated that kernelization provides little to no reduction in size for these instances.
Therefore, we omit them from further evaluation and focus on more sparse graphs.

\begin{table}[b!]
  \caption{Impact of kernelization on the computation of a maximum cut by \textsc{LocalSolver}~(LS) and \textsc{Biq Mac} (BM).
    {Times are given in seconds.
    Kernelization is accounted for within the timings for $G_{\textnormal{ker}}$.}
    Values in brackets provide the speedup and are derived from $\frac{T(G)}{T(G_{\textnormal{ker}})}$.
    Times labeled with~``\unaryminus'' exceeded the ten-hour time limit and an ``f'' indicates the solver crashed. \label{solver:weighted}}
  \footnotesize
  \begin{center}
    \setlength{\tabcolsep}{.7ex}
    \begin{tabular}{l@{\hskip 10pt}rr@{\hskip 25pt} rrr@{\hskip 15pt}rrr}
      \toprule
      \multicolumn{1}{l}{{Name}} 
       
        & \multicolumn{1}{c}{{${|V|}$}}
        
        & \multicolumn{1}{l}{{${e(G)}$}}
        & \multicolumn{1}{c}{{${T_{\textnormal{LS}}(G)}$}}
        & \multicolumn{2}{c}{{${T_{\textnormal{LS}}(G_{\textnormal{ker}})}$}}
        
        & \multicolumn{1}{c}{{${T_{\textnormal{BM}}(G)}$}}
        & \multicolumn{2}{c}{{${T_{\textnormal{BM}}(G_{\textnormal{ker}})}$}}
        \\
               \midrule 
        \Id{\detokenize{ca-CSphd}}         & \numprint{1882} & 0.99 & 24.07              & 0.32   & [75.40]    & -      & 0.06                & [$\infty$] \\
        \Id{\detokenize{ego-facebook}}     & \numprint{2888} & 1.00 & 20.09              & 0.09   & [228.91]   & -      & 0.01                & [$\infty$] \\
        \Id{\detokenize{ENZYMES_g295}}    & 123             & 0.86 & 1.22               & 0.33   & [3.70]     & 0.82   & 0.13                & [6.57] \\
        \Id{\detokenize{road-euroroad}}    & \numprint{1174} & 0.79 & -                  & -      & -          & -      & -                   & - \\
        \Id{\detokenize{bio-yeast}}        & 1458            & 0.81 & -                  & -      & -          & -      & \numprint{32726.75} & [$\infty$] \\
        \Id{\detokenize{rt-twitter-copen}} & 761             & 0.85 & -                  & 834.71 & [$\infty$] & -      & 1.77                & [$\infty$] \\
        \Id{\detokenize{bio-diseasome}}    & 516             & 0.93 & -                  & 4.91   & [$\infty$] & -      & 0.07                & [$\infty$] \\
        \Id{\detokenize{ca-netscience}}    & 379             & 0.77 & -                  & 956.03 & [$\infty$] & -      & 0.67                & [$\infty$] \\
        \Id{\detokenize{soc-firm-hi-tech}} & 33              & 0.36 & 4.67               & 1.61   & [2.90]     & 0.09   & 0.06                & [1.41] \\
        \Id{\detokenize{g000302}}          & 317             & 0.21 & 0.58               & 0.49   & [1.17]     & 1.88   & 0.74                & [2.53] \\
        \Id{\detokenize{g001918}}          & 777             & 0.12 & 1.47               & 1.41   & [1.04]     & 31.11  & 17.45               & [1.78] \\
        \Id{\detokenize{g000981}}          & 110             & 0.28 & 10.73              & 4.73   & [2.27]     & 531.47 & 21.53               & [24.68] \\
        \Id{\detokenize{g001207}}          & 84              & 0.19 & 1.10               & 0.16   & [6.88]     & 53.20  & 0.06                & [962.38] \\
        \Id{\detokenize{g000292}}          & 212             & 0.03 & 0.45               & 0.45   & [1.01]     & 0.43   & 0.37                & [1.14] \\
        \Id{\detokenize{imgseg_271031}}    & 900             & 0.99 & 10.66              & 0.19   & [55.94]    & -      & 0.17                & [$\infty$] \\
        \Id{\detokenize{imgseg_105019}}    & \numprint{3548} & 0.93 & 234.01             & 22.68  & [10.32]    & f      & \numprint{13748.62} & [$\infty$] \\
        \Id{\detokenize{imgseg_35058}}     & \numprint{1274} & 0.37 & 34.93              & 24.71  & [1.41]     & -      & -                   & - \\
        \Id{\detokenize{imgseg_374020}}    & \numprint{5735} & 0.82 & \numprint{1739.11} & 72.23  & [24.08]    & f      & -                   & - \\
        \Id{\detokenize{imgseg_106025}}    & \numprint{1565} & 0.68 & 159.31             & 34.05  & [4.68]     & -      & -                   & - \\
        \bottomrule
    \end{tabular}
  \end{center}
\end{table}
\begin{table}[t]
  \caption{Impact of kernelization on the computation of a maximum cut by \textsc{LocalSolver}~(LS) and \textsc{Biq Mac} (BM). {Times are given in seconds. Kernelization time is included in the solving times for $G_{\textnormal{ker}}$.} Values in brackets provide the speedup and are derived from $\frac{T(G)}{T(G_{\textnormal{ker}})}$. Times labeled with~``\unaryminus'' exceeded the ten-hour time limit. Weighted path compression by Reduction Rule~\ref{r:weighted:compression} \emph{is not} used at the end -- the kernel is unweighted.\label{solver:unweighted}}
  \footnotesize
  \begin{center}
    \setlength{\tabcolsep}{.7ex}
    \begin{tabular}{l@{\hskip 10pt}rr@{\hskip 25pt} rrr@{\hskip 15pt}rrr}
      \toprule
    \multicolumn{1}{l}{{Name}} 
    
    & \multicolumn{1}{c}{{${|V|}$}}
    
    & \multicolumn{1}{l}{{${e(G)}$}}
    & \multicolumn{1}{c}{{${T_{\textnormal{LS}}(G)}$}}
    & \multicolumn{2}{c}{{${T_{\textnormal{LS}}(G_{\textnormal{ker}})}$}}
    
    & \multicolumn{1}{c}{{${T_{\textnormal{BM}}(G)}$}}
    & \multicolumn{2}{c}{{${T_{\textnormal{BM}}(G_{\textnormal{ker}})}$}}
    \\ \midrule

     \Id{\detokenize{ca-CSphd}}        & \numprint{1882} & 0.98 & 24.79 & 1.12    & [22.23]    & -      & 0.32   & [$\infty$] \\
     \Id{\detokenize{ego-facebook}}     & \numprint{2888} & 0.93 & 20.39 & 1.72    & [11.83]    & 967.99 & 1.42   & [682.04] \\
     \Id{\detokenize{ENZYMES_g295}}   & 123  & 0.82 & 1.83  & 0.36    & [5.09]     & 0.96   & 0.37   & [2.60] \\
     \Id{\detokenize{road-euroroad}}    &\numprint{1174} & 0.69 & -     & -       & -          & -      & -      & - \\
     \Id{\detokenize{bio-yeast}}        & \numprint{1458} & 0.72 & -     & -       & -          & -      & -      & - \\
     \Id{\detokenize{rt-twitter-copen}} & 761  & 0.80 & -     & 409.47  & [$\infty$] & -      & 101.14 & [$\infty$] \\
     \Id{\detokenize{bio-diseasome}}    & 516  & 0.93 & -     & 6.66    & [$\infty$] & -      & 0.35   & [$\infty$] \\
     \Id{\detokenize{ca-netscience}}    & 379  & 0.67 & -     & \numprint{4116.61} & [$\infty$] & -      & 2.10   & [$\infty$] \\
     \Id{\detokenize{soc-firm-hi-tech}} & 33   & 0.30 & 4.92  & 2.34    & [2.10]     & 0.29   & 0.31   & [0.94] \\
     \Id{\detokenize{g000302}}          & 317  & 0.10 & 0.71  & 0.50    & [1.41]     & 1.28   & 0.89   & [1.44] \\
     \Id{\detokenize{g001918}}          & 777  & 0.06 & 1.67  & 1.51    & [1.10]     & 14.90  & 11.69  & [1.27] \\
     \Id{\detokenize{g000981}}          & 110  & 0.22 & 11.32 & 1.97    & [5.74]     & 0.98   & 0.44   & [2.23] \\
     \Id{\detokenize{g001207}}          & 84   & 0.17 & 1.56  & 0.15    & [10.11]    & 0.47   & 0.37   & [1.28] \\
     \Id{\detokenize{g000292}}          & 212  & 0.01 & 0.69  & 0.51    & [1.35]     & 0.56   & 0.62   & [0.91] \\
    \bottomrule
   \end{tabular}
  \end{center}
\end{table}

First, we notice that kernelization is able to provide moderate to significant speedups for all instances that we have tested.
In particular, we are able to a speedup between $1.04$ and $228.91$ for instances that were previously solvable by \textsc{LocalSolver}.
Likewise, for the instances that \textsc{Biq Mac} is able to process, we achieve a speedup of up to three orders of magnitude.
Furthermore, we allow these solvers to now compute a maximum cut for a majority of instances that have previously been infeasible in less than $17$ minutes.

To examine the impact when allowing a weighted kernel, we now compare the performance our algorithm using weighted path compression (Table~\ref{solver:weighted}) with the unweighted version (Table~\ref{solver:unweighted}).
We can see that by including weighted path compression we can achieve significantly better speedups, especially for the sparse real-world instances by Rossi and Ahmed~\mycite{nr}.
For example, on \texttt{ego-facebook} we achieve a speedup of $228.91$ with compression and $11.83$ without.

Finally, it is also noteworthy that we get significant improvements for the weighted instances from VLSI design and image segmentation.
By examining the performance of each individual reduction rule, we can see that this is solely due to Reduction Rule~\ref{rule:x1:scaled}.
These findings could improve the work by de Sousa et al.~\mycite{de2013estimation}, which also affects the work by Dunning et al.~\mycite{dunning-2018}.
In conclusion, our novel reduction rules give us a simple but powerful tool for speeding up existing state-of-the-art solvers for computing maximum cuts.
Moreover, as mentioned previously, even our simple weighted path compression by itself is able to have a significant impact.

\subsection{Analysis on Large Instances}
\label{subsec:LargeEvaluation}
We now examine the performance of our kernelization framework and its impact on existing solvers for large graph instances with up to millions of vertices.
For this purpose, we compared the cut size over time achieved by \textsc{LocalSolver} and \textsc{MqLib} with and without our kernelization. 
Note that we did not use \textsc{Biq Mac} as it was not able to handle instances with more than \numprint{3000} vertices.
Our results using a three-hour time limit for each solver are given in Table~\ref{eval:big}.
Furthermore, we present convergence plots in Fig.~\ref{convergence:big}.
\begin{table}[htb!]
  \caption{Evaluation of large graph instances.
  A three-hour time limit was used and five iterations were performed.
  The columns $\Delta_\textnormal{LS}$ and $\Delta_\textnormal{MQ}$ indicate the percentage by which the size of the largest computed cut is larger on the kernelized graph compared to the non-kernelized one, for \textsc{LocalSolver} and \textsc{MqLib}, respectively.\label{eval:big}}
  \footnotesize
  \begin{center}
    \begin{tabular}{l@{\hskip 10pt}rr@{\hskip 25pt} rrr@{\hskip 15pt}rrr}
      \toprule
      \multicolumn{1}{l}{Name} 
      & \multicolumn{1}{c}{\textbf{$|V|$}}
      & \multicolumn{1}{l}{\textbf{$deg_{\textnormal{avg}}$}} 
      
      & \multicolumn{1}{c}{$e(G)$}
      & \multicolumn{1}{c}{$T_{\textnormal{ker}}(G)$}
      & \multicolumn{1}{c}{$\Delta_\textnormal{LS}$}
      & \multicolumn{1}{c}{$\Delta_\textnormal{MQ}$}\\ \midrule
      \Id{\detokenize{inf-road_central}} & \numprint{14081816}    &  1.20  & 0.59 & 362.32 & inf\% & 2.70\%  \\ 
      \Id{\detokenize{inf-power}}         &     \numprint{4941}    &  1.33  & 0.62 &  0.04 &  1.64\% & 0.45\%  \\ 
      \Id{\detokenize{web-google}}        &    \numprint{1299}    &  2.13 & 0.79 &  0.01 &  0.69\% & 0.19\%  \\ 
      \Id{\detokenize{ca-MathSciNet}}     &   \numprint{332689}    &  2.47  & 0.63 &  8.02 &  1.33\% & 0.55\%   \\ 
      \Id{\detokenize{ca-IMDB}}    	    &   \numprint{896305}    &  4.22  & 0.42 & 27.55 &  0.97\% & 0.32\%        \\ 
      \Id{\detokenize{web-Stanford}}      &   \numprint{281903}    &  7.07  & 0.18 & 105.17 &  0.34\% & 0.30\%  \\ 
      \Id{\detokenize{web-it-2004}}       &   \numprint{509338}    & 14.09  & 0.91 & 22.10 &  0.08\% & 0.02\%  \\ 
      \Id{\detokenize{ca-coauthors-dblp}} &   \numprint{540486}    & 28.20  & 0.25 & 72.39  &  0.05\% & 0.04\% \\ 
     \bottomrule 
    \end{tabular}
  \end{center}
\end{table}

First, we note that the time to compute the actual kernel is relatively small. 
In particular, we are able to compute a kernel for a graph with $14$ million vertices and edges in just over six minutes.
Furthermore, we achieve an efficiency between $0.18$ and $0.91$ across all tested instances.
When looking at the convergence plots (Fig.~\ref{convergence:big}) we can observe that the additional preprocessing time of kernelization is quickly compensated by a significantly steeper increase in cut size compared to the unkernelized version.
Furthermore, for instances where a kernel can be computed very quickly, such as \texttt{web-google}, we find a better solution almost instantaneously.
In general, the results achieved by kernelization followed by the local search heuristic are \emph{always} better than just using the local search heuristic alone.
However, the final improvement on the size of the largest cut found by \textsc{LocalSolver} and \textsc{MqLib} is generally small for the given time limit of three hours.
\begin{figure}[htb!]
  \centering
  \includegraphics[width=0.95\textwidth]{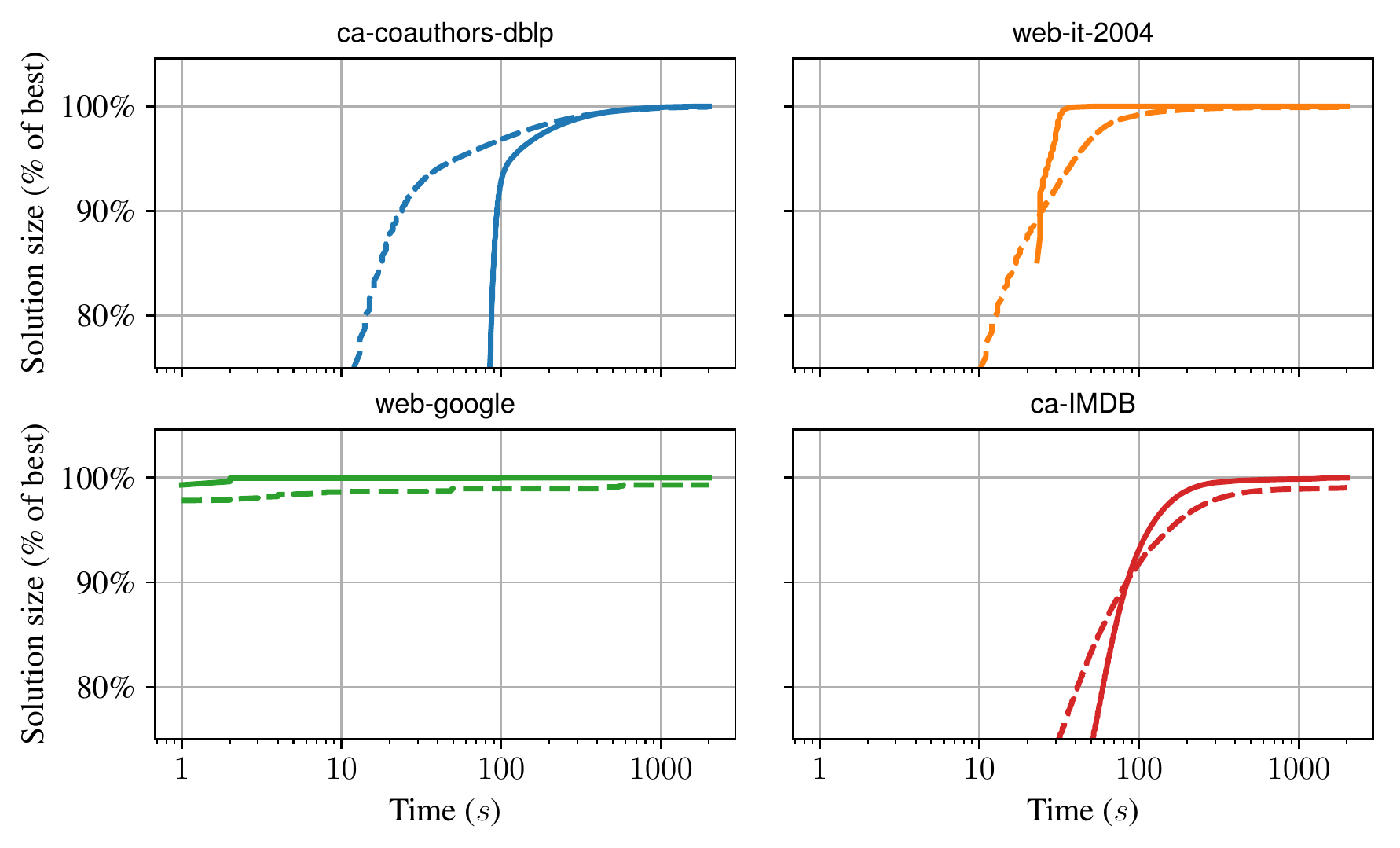}
  \caption{Convergence of \textsc{LocalSolver} on large instances. The dashed line represents the size of the cut for the non-kernelized graph, while the full line does so for the kernelized graph.\label{convergence:big}}
\end{figure}

\section{Conclusions}
\label{sec:Conclusion}
We engineered new efficient data reduction rules for \textsc{Max Cut} and showed that these rules subsume most existing rules.
Our extensive experiments show that kernelization has a significant impact in practice.
In particular, our experiments reveal that current state-of-the-art solvers can be sped up by up to \emph{multiple orders of magnitude} when combined with our data reduction rules.

Developing new reduction rules is an important direction for future research.
Of particular interest are reduction rules for \textsc{Weighted Max Cut}, where reduction rules yield a weighted kernel.

\vfill
\pagebreak
\bibliography{references}

\begin{thebibliography}{10}

\bibitem{biqmac}
{BiqMac Library}.
\newblock \url{http://biqmac.aau.at/biqmaclib.html}, 2018.
\newblock [Online; accessed 2-September-2018].

\bibitem{abu2007kernelization}
Faisal~N. Abu{-}Khzam, Michael~R. Fellows, Michael~A. Langston, and W.~Henry
  Suters.
\newblock Crown structures for vertex cover kernelization.
\newblock {\em Theory Comput. Syst.}, 41(3):411--430, 2007.
\newblock \href {http://dx.doi.org/10.1007/s00224-007-1328-0}
  {\path{doi:10.1007/s00224-007-1328-0}}.

\bibitem{Arraiz2009}
Emely Arr\'{a}iz and Oswaldo Olivo.
\newblock Competitive simulated annealing and tabu search algorithms for the
  {M}ax-{C}ut problem.
\newblock In {\em Proceedings of the 11th Annual Conference on Genetic and
  Evolutionary Computation}, GECCO '09, pages 1797--1798, New York, NY, USA,
  2009. ACM.
\newblock \href {http://dx.doi.org/10.1145/1569901.1570167}
  {\path{doi:10.1145/1569901.1570167}}.

\bibitem{barahona1982computational}
Francisco Barahona.
\newblock On the computational complexity of {I}sing spin glass models.
\newblock {\em J. Phys. A: Mathematical and General}, 15(10):3241, 1982.
\newblock \href {http://dx.doi.org/10.1088/0305-4470/15/10/028}
  {\path{doi:10.1088/0305-4470/15/10/028}}.

\bibitem{barahona1996network}
Francisco Barahona.
\newblock Network design using cut inequalities.
\newblock {\em SIAM J. Optim.}, 6(3):823--837, 1996.
\newblock \href {http://dx.doi.org/10.1137/S1052623494279134}
  {\path{doi:10.1137/S1052623494279134}}.

\bibitem{barahona1988application}
Francisco Barahona, Martin Gr{\"o}tschel, Michael J{\"u}nger, and Gerhard
  Reinelt.
\newblock An application of combinatorial optimization to statistical physics
  and circuit layout design.
\newblock {\em Oper. Res.}, 36(3):493--513, 1988.
\newblock \href {http://dx.doi.org/10.1287/opre.36.3.493}
  {\path{doi:10.1287/opre.36.3.493}}.

\bibitem{Benlic2013}
Una Benlic and Jin-Kao Hao.
\newblock Breakout local search for the {M}ax-{C}ut problem.
\newblock {\em Engineering Applications of Artificial Intelligence},
  26(3):1162--1173, 2013.
\newblock \href {http://dx.doi.org/10.1016/j.engappai.2012.09.001}
  {\path{doi:10.1016/j.engappai.2012.09.001}}.

\bibitem{localsolverx1}
Thierry Benoist, Bertrand Estellon, Fr{\'e}d{\'e}ric Gardi, Romain Megel, and
  Karim Nouioua.
\newblock Localsolver 1.x: a black-box local-search solver for 0-1 programming.
\newblock {\em 4OR}, 9(3):299, 2011.
\newblock [used in this work: Localsolver 8.0].
\newblock URL: \url{https://www.localsolver.com/}, \href
  {http://dx.doi.org/10.1007/s10288-011-0165-9}
  {\path{doi:10.1007/s10288-011-0165-9}}.

\bibitem{chiang2007fast}
Charles Chiang, Andrew~B Kahng, Subarnarekha Sinha, Xu~Xu, and Alexander~Z
  Zelikovsky.
\newblock Fast and efficient bright-field {AAPSM} conflict detection and
  correction.
\newblock {\em IEEE Transactions on Computer-Aided Design of Integrated
  Circuits and Systems}, 26(1):115--126, 2007.
\newblock \href {http://dx.doi.org/10.1109/TCAD.2006.882642}
  {\path{doi:10.1109/TCAD.2006.882642}}.

\bibitem{DBLP:journals/corr/abs-1212-6848}
Robert Crowston, Gregory Gutin, Mark Jones, and Gabriele Muciaccia.
\newblock Maximum balanced subgraph problem parameterized above lower bound.
\newblock {\em Theoret. Comput. Sci.}, 513:53--64, 2013.
\newblock \href {http://dx.doi.org/10.1016/j.tcs.2013.10.026}
  {\path{doi:10.1016/j.tcs.2013.10.026}}.

\bibitem{DBLP:journals/corr/abs-1112-3506}
Robert Crowston, Mark Jones, and Matthias Mnich.
\newblock Max-{C}ut parameterized above the {E}dwards-{E}rd{\H{o}}s bound.
\newblock {\em Algorithmica}, 72(3):734--757, 2015.
\newblock \href {http://dx.doi.org/10.1007/s00453-014-9870-z}
  {\path{doi:10.1007/s00453-014-9870-z}}.

\bibitem{de1959file}
Rene De~La~Briandais.
\newblock File searching using variable length keys.
\newblock In {\em Papers presented at the the March 3-5, 1959, western joint
  computer conference}, pages 295--298. ACM, 1959.
\newblock \href {http://dx.doi.org/10.1145/1457838.1457895}
  {\path{doi:10.1145/1457838.1457895}}.

\bibitem{de2013estimation}
Samuel de~Sousa, Yll Haxhimusa, and Walter~G Kropatsch.
\newblock Estimation of distribution algorithm for the max-cut problem.
\newblock In {\em International Workshop on Graph-Based Representations in
  Pattern Recognition}, volume 7877 of {\em LNCS}, pages 244--253. Springer,
  2013.
\newblock \href {http://dx.doi.org/10.1007/978-3-642-38221-5_26}
  {\path{doi:10.1007/978-3-642-38221-5_26}}.

\bibitem{dunning-2018}
Iain Dunning, Swati Gupta, and John Silberholz.
\newblock What works best when? {A} systematic evaluation of heuristics for
  {Max-Cut} and {QUBO}.
\newblock {\em INFORMS J. Comput.}, 30(3):608--624, 2018.
\newblock \href {http://dx.doi.org/10.1287/ijoc.2017.0798}
  {\path{doi:10.1287/ijoc.2017.0798}}.

\bibitem{edwards1973some}
Christopher~S Edwards.
\newblock Some extremal properties of bipartite subgraphs.
\newblock {\em Canad. J. Math.}, 25(3):475--485, 1973.
\newblock \href {http://dx.doi.org/10.4153/CJM-1973-048-x}
  {\path{doi:10.4153/CJM-1973-048-x}}.

\bibitem{edwards1975improved}
Christopher~S Edwards.
\newblock An improved lower bound for the number of edges in a largest
  bipartite subgraph.
\newblock In {\em Proc. Second Czechoslovak Symposium on Graph Theory, Prague},
  pages 167--181, 1975.

\bibitem{Etscheid2018}
Michael Etscheid and Matthias Mnich.
\newblock Linear kernels and linear-time algorithms for finding large cuts.
\newblock {\em Algorithmica}, 80(9):2574--2615, 2018.
\newblock \href {http://dx.doi.org/10.1007/s00453-017-0388-z}
  {\path{doi:10.1007/s00453-017-0388-z}}.

\bibitem{DBLP:journals/corr/FariaKSS15}
Luerbio Faria, Sulamita Klein, Ignasi Sau, and Rubens Sucupira.
\newblock Improved kernels for signed max cut parameterized above lower bound
  on ($r$, $l$)-graphs.
\newblock {\em Discrete Math. \& Theoret. Comput. Sci.}, 19(1), 2017.
\newblock \href {http://dx.doi.org/10.23638/DMTCS-19-1-14}
  {\path{doi:10.23638/DMTCS-19-1-14}}.

\bibitem{maFerizovic}
Damir Ferizovic.
\newblock {A Practical Analysis of Kernelization Techniques for the Maximum Cut
  Problem}.
\newblock {Master's Thesis}, Karlsruhe Institute of Technology, 2019.

\bibitem{fredkin1960trie}
Edward Fredkin.
\newblock Trie memory.
\newblock {\em Comm. ACM}, 3(9):490--499, 1960.
\newblock \href {http://dx.doi.org/10.1145/367390.367400}
  {\path{doi:10.1145/367390.367400}}.

\bibitem{funke2019communication}
Daniel Funke, Sebastian Lamm, Ulrich Meyer, Manuel Penschuck, Peter Sanders,
  Christian Schulz, Darren Strash, and Moritz von Looz.
\newblock Communication-free massively distributed graph generation.
\newblock {\em Journal of Parallel and Distributed Computing}, 131:200--217,
  2019.
\newblock \href {http://dx.doi.org/10.1016/j.jpdc.2019.03.011}
  {\path{doi:10.1016/j.jpdc.2019.03.011}}.

\bibitem{localsolverx2}
Fr{\'e}d{\'e}ric Gardi, Thierry Benoist, Julien Darlay, Bertrand Estellon, and
  Romain Megel.
\newblock {\em Mathematical Programming Solver Based on Local Search}.
\newblock FOCUS Series in Computer Engineering. ISTE Wiley, 2014.
\newblock \href {http://dx.doi.org/10.1002/9781118966464}
  {\path{doi:10.1002/9781118966464}}.

\bibitem{harary1959measurement}
Frank Harary.
\newblock On the measurement of structural balance.
\newblock {\em Behavioral Sci.}, 4(4):316--323, 1959.
\newblock \href {http://dx.doi.org/10.1002/bs.3830040405}
  {\path{doi:10.1002/bs.3830040405}}.

\bibitem{harary2002signed}
Frank Harary, Meng-Hiot Lim, and Donald~C Wunsch.
\newblock Signed graphs for portfolio analysis in risk management.
\newblock {\em IMA J. Mgmt. Math.}, 13(3):201--210, 2002.
\newblock \href {http://dx.doi.org/10.1093/imaman/13.3.201}
  {\path{doi:10.1093/imaman/13.3.201}}.

\bibitem{hespe2017scalable}
Demian Hespe, Christian Schulz, and Darren Strash.
\newblock Scalable kernelization for maximum independent sets.
\newblock In {\em Proc. ALENEX 2018}, pages 223--237, 2018.
\newblock \href {http://dx.doi.org/10.1137/1.9781611975055.19}
  {\path{doi:10.1137/1.9781611975055.19}}.

\bibitem{karp1972reducibility}
Richard~M Karp.
\newblock Reducibility among combinatorial problems.
\newblock In {\em Complexity of Computer Computations}, The IBM Research
  Symposia Series, pages 85--103. Springer, 1972.
\newblock \href {http://dx.doi.org/10.1007/978-1-4684-2001-2_9}
  {\path{doi:10.1007/978-1-4684-2001-2_9}}.

\bibitem{Kochenberger2013}
Gary~A. Kochenberger, Jin-Kao Hao, Zhipeng L{\"u}, Haibo Wang, and Fred Glover.
\newblock Solving large scale {Max Cut} problems via tabu search.
\newblock {\em Journal of Heuristics}, 19(4):565--571, Aug 2013.
\newblock \href {http://dx.doi.org/10.1007/s10732-011-9189-8}
  {\path{doi:10.1007/s10732-011-9189-8}}.

\bibitem{lamm2017finding}
Sebastian Lamm, Peter Sanders, Christian Schulz, Darren Strash, and Renato~F
  Werneck.
\newblock Finding near-optimal independent sets at scale.
\newblock {\em J. Heuristics}, 23(4):207--229, 2017.
\newblock \href {http://dx.doi.org/10.1007/s10732-017-9337-x}
  {\path{doi:10.1007/s10732-017-9337-x}}.

\bibitem{DBLP:conf/csr/Madathil0Z18}
Jayakrishnan Madathil, Saket Saurabh, and Meirav Zehavi.
\newblock Max-{C}ut {A}bove {S}panning {T}ree is fixed-parameter tractable.
\newblock In {\em Proc. CSR 2018}, volume 10846 of {\em LNCS}, pages 244--256.
  Springer, 2018.

\bibitem{Prieto:2005:MES:1082260.1082274}
Elena Prieto.
\newblock The method of extremal structure on the $k$-{M}aximum {C}ut problem.
\newblock In {\em Proc. CATS 2005}, volume~41, pages 119--126. ACM, 2005.

\bibitem{RRW10}
Franz Rendl, Giovanni Rinaldi, and Angelika Wiegele.
\newblock Solving {M}ax-{C}ut to optimality by intersecting semidefinite and
  polyhedral relaxations.
\newblock {\em Math. Prog.}, 121(2):307, 2010.
\newblock \href {http://dx.doi.org/10.1007/s10107-008-0235-8}
  {\path{doi:10.1007/s10107-008-0235-8}}.

\bibitem{nr}
Ryan~A Rossi and Nesreen~K Ahmed.
\newblock The network data repository with interactive graph analytics and
  visualization.
\newblock In {\em Proc. AAAI 2015}, volume~15, pages 4292--4293, 2015.
\newblock URL:
  \url{https://www.aaai.org/ocs/index.php/AAAI/AAAI15/paper/view/9553/9856}.

\bibitem{sanders2016generators}
Peter Sanders and Christian Schulz.
\newblock Scalable generation of scale-free graphs.
\newblock {\em Inf. Proc. Lett.}, 116(7):489--491, 2016.
\newblock \href {http://dx.doi.org/10.1016/j.ipl.2016.02.004}
  {\path{doi:10.1016/j.ipl.2016.02.004}}.

\bibitem{van1993linear}
Nguyen Van~Ngoc and Zsolt Tuza.
\newblock Linear-time approximation algorithms for the max cut problem.
\newblock {\em Combinatorics, Probability Comput.}, 2(2):201--210, 1993.
\newblock \href {http://dx.doi.org/10.1017/S0963548300000596}
  {\path{doi:10.1017/S0963548300000596}}.

\bibitem{Wang2013}
Yang Wang and Zhipeng L\.
\newblock Probabilistic {GRASP}-tabu search algorithms for the {UBQP} problem.
\newblock {\em Computers \& Operations Research}, 40(12):3100--3107, 2013.
\newblock \href {http://dx.doi.org/10.1016/j.cor.2011.12.006}
  {\path{doi:10.1016/j.cor.2011.12.006}}.

\end{thebibliography}

\end{document}